\documentclass{svproc}
\usepackage{xcolor}
\usepackage{amsmath,amssymb}
\usepackage{epsfig}
\usepackage{amssymb}
\usepackage{graphicx,float}
\usepackage{subcaption}
\usepackage{url}

\date{}
\begin{document}

\mainmatter

\title{Winner does not take all: contrasting centrality in adversarial networks}

\author{Anthony Bonato\inst{1}\thanks{Supported by an NSERC Discovery Grant.} \and Joey Kapusin\inst{1} \and Jiajie Yuan\inst{1}}

\institute{Toronto Metropolitan University, Toronto, Ontario, Canada}

\maketitle

\begin{abstract}
In adversarial networks, edges correspond to negative interactions such as competition or dominance. We introduce a new type of node called a low-key leader in adversarial networks, distinguished by contrasting the centrality measures of CON score and PageRank. We present a novel hypothesis that low-key leaders are ubiquitous in adversarial networks and provide evidence by considering data from real-world networks, including dominance networks in 172 animal populations, trading networks between G20 nations, and Bitcoin trust networks. We introduce a random graph model that generates directed graphs with low-key leaders. 
\end{abstract}

\section{Introduction}\label{intro}

Adversarial networks, where edges capture competition, dominance, or enmity, are gaining prominence in the study of complex networks. Negative interactions are critically important to the study of social networks and more broadly, real-world complex networks, and are often hidden
drivers of link formation. Adversarial networks appear throughout network science, and examples range from negatively correlated stocks in market graphs \cite{bog}, trade between nations \cite{li}, the spatial location of cities as a model to predict the rise of conflicts and violence \cite{guo}, and animal predation networks and food webs \cite{food}. Even in the highly cited Zachary Karate club network \cite{zach}, the negative interaction between the administrator and instructor was the impetus for the split of the club participants into two communities. 

The \emph{Dynamic Competition Hypothesis} (or DCH) was introduced in \cite{DCH} and provides a quantitative framework for the structure of evolving adversarial or competition networks. The DCH posits that leaders in adversarial networks exhibit high closeness, low in-degree, and high out-degree. Leaders also possess high common out-neighbor (or CON) scores, which measures shared competition versus other nodes; see Section~\ref{secCON} for a definition and discussion of CON scores. 

We focus on the role of competing notions of centrality in adversarial networks. Centrality measures are used to identify certain key nodes within complex networks. There are several methods to measure centrality, such as degree distribution, PageRank, closeness, and betweenness. Our focus will be on the novel detection and analysis of certain nodes in adversarial networks, measured by comparing CON scores and PageRank; the latter measure is an established tool for determining influential nodes in a network.

In Section~\ref{secCON}, we identify a low-key leader as a node which has a relatively high CON score, but low PageRank; intuitively, a low-key leader is highly likely to affect link evolution while remaining less visible in the network. This notion is analogous to so-called silent or quiet leaders in management positions in companies, who may be more diplomatic, introverted, but remain influential; see \cite{grant}. Low-key leaders appear to be ubiquitous in adversarial networks, and we support this hypothesis in Section~\ref{datas} with data from three distinct sources: dominance networks in 172 distinct animal populations, trading networks between G20 nations, and Bitcoin trust networks. A new random graph model is introduced in Section~3, with the aim of synthetically generating low-key leaders in scale-free directed graphs. The concluding section contains several directions for future research.

We consider directed graphs (or \emph{digraphs}) with multiple directed edges in the paper. Additional background on graph theory and complex networks may be found in the book \cite{west}.

\section{Low-key Leaders}\label{secCON}

Survivor is a popular social game television franchise, where contestants progressively eliminate each other by voting until only one remains. In the 35th season of the American social game show Survivor, Ben Driebergen won over finalists Chrissy Hofbeck and Ryan Ulrich \cite{wiki}. While Ryan and Chrissy played a strategic game throughout the reality show competition by forging alliances and voting out key competitors, Ben won in part based on his finding multiple immunity idols and the sympathy he garnered from the jury as a veteran. Although Ben won the game, more low-key players like Ryan were instrumental in shaping the underlying adversarial, co-voting network. We quantify this phenomenon of low-key but influential actors in networks via centrality measures in adversarial networks.

An approach taken in \cite{DCH,centrality} in the detection of leaders in adversarial networks is the common out-neighbour score (or CON score).  For nodes $u, v, w$ in a graph $G$, we define $w$ to be a \emph{common out-neighbor} of $u$ and $v$ if ($u,w$) and ($v, w$) are two directed edges in $G$. We let CON($u,v$) be the number of common out-neighbour of distinct nodes $u$ and $v$, and define $$\mathrm{CON}(u)=\sum_{v\in V(G)} \mathrm{CON}(u,v).$$ 
A high CON score for a node indicates it shares many of the same adversaries with other nodes, and hence, is more in sync with how links evolve in the network. A low CON score indicates the opposite trait, where the node is less of a driver of link evolution.

PageRank centrality is based on the stationary distribution of a random walk on the network that periodically teleports to a node chosen uniformly at random. In adversarial networks, we compute the PageRank of nodes on the reversed-edge network, where we change the orientation of the directed edges. Hence, if a node in the network has many out-edges, they will more likely have higher PageRank in the reversed-edge network.  

We define a \emph{low-key leader} (or \emph{LKL}) in an adversarial networks as a node whose CON score and PageRank are negatively correlated, with high CON score and low PageRank. Recall that, according to the DCH, leaders in a network are nodes that exhibit high closeness, high CON score, low in-degree, and high out-degree. In contrast, low-key leaders have less centrality due to their low PageRank but remain influential actors in the network owing to their high CON score.  

The definition of low-key leader given in the previous paragraph is more heuristic, as having low or high scores is subject to interpretation. To make the definition more precise, we consider the following approach. While CON scores are integers, PageRank consists of probabilities in $[0,1]$. To compare the difference between the two scores to validate the presence of LKLs, we re-scale both scores by using the \emph{unity-based normalization}, defined as follows. Suppose we are given real numbers $X_1,X_2,\ldots ,X_n$, with minimum $X_{\mathrm{min}}$ and maximum $X_{\mathrm{max}}$. For $1\le i \le n,$ define
$$X_{i,\mathrm{norm}}=\frac{X_i-X_{\mathrm{min}}}{X_{\mathrm{max}}-X_{\mathrm{min}}}.$$
Such scaling measure is used to set all values $X_{i,\mathrm{norm}} \in [0,1]$; note that we apply this normalization also to PageRank (whose values are already in $[0,1]$) for consistency. 

Suppose that for a set of nodes $x_i,$ where $1\le i\le n,$ the CON score and PageRank of $x_i$ are denoted by CON$_i$ and PR$_i$, respectively. Define $$\varepsilon_i = \mathrm{CON}_{i,\mathrm{norm}} - \mathrm{PR}_{i,\mathrm{norm}}.$$ Note that $\varepsilon \in [-1,1].$ We abuse notation and refer to $\varepsilon_i$ as simply $\varepsilon.$ We consider a node to be a low-key leader if it has the maximum value of $\varepsilon,$ and $\varepsilon > 0.5.$ We refer to $\varepsilon$ as the \emph{low-key leader strength} of a node.

We hypothesize that adversarial networks typically contain at least one low-key leader. Note that the assertion is on the presence of influential nodes within adversarial networks; no other data is required other than the presence of negative ties. We provide evidence for the hypothesis in real-world, adversarial networks in the next section.

\section{Data and Methods}\label{datas}

To validate our hypothesis on low-key leaders, we consider three types of adversarial networks: dominance networks in animal groups, Bitcoin trust networks, and the trading networks between nations. In the interest of space, we refer the reader to \url{https://github.com/jkapusin/Low-Key-Leaders} for complete data sets, as well as CON, PageRank, and low-key leader strengths of nodes for each network. 

\subsection{Dominance Networks}

We first consider dominance networks, where directed edges correspond to some form of dominant-subordinate relations between members of an animal population. The animal social dominance data set, compiled by Shizuka and McDonald \cite{dyrad} and available in the Dryad Digital Repository, contains 172 distinct dominance networks of an animal group. Networks are represented as weighted adjacency matrices, and each entry in a given matrix corresponds to the number of times that the animal in the row is dominated by the one in the column. 

\begin{figure}[h!]
\centering
\includegraphics[scale=0.23]{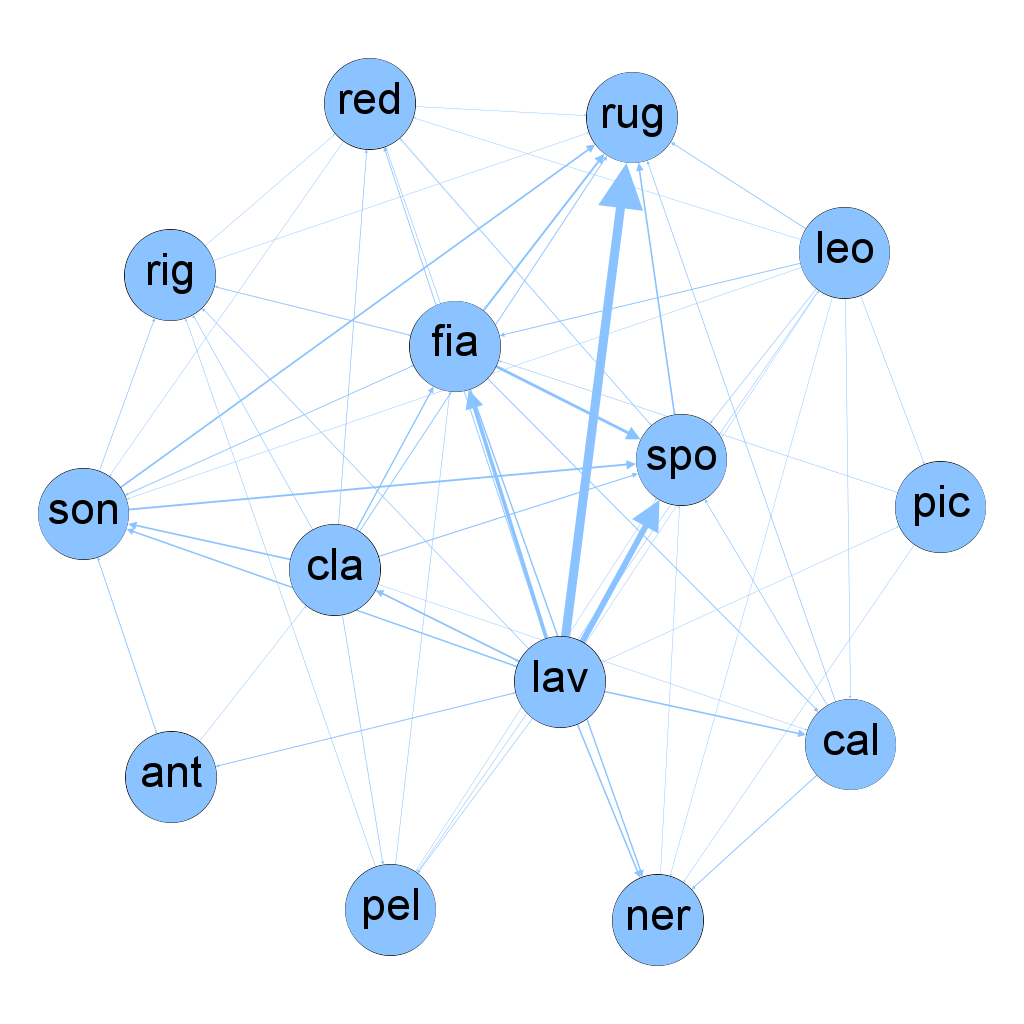}
\caption{The Bonanni2007-2 dominance network.}\label{bonfig11}
\end{figure}

We first consider more closely the Bonanni2007-2 data set, which was randomly chosen. The Bonanni2007-2 data set contains information on dominance behaviour of mongrel dogs living in a free-ranging or semi-free-ranging state. Directed edges correspond to aggressive signals such as lunging, biting, or snarling between the dogs. Each individual dog is identified using a three letter code. See Figure~\ref{bonfig11} for a visualization of this network and see \cite{bonanni} for more discussion.

\begin{figure}[h!]
\centering
\includegraphics[scale=0.26]{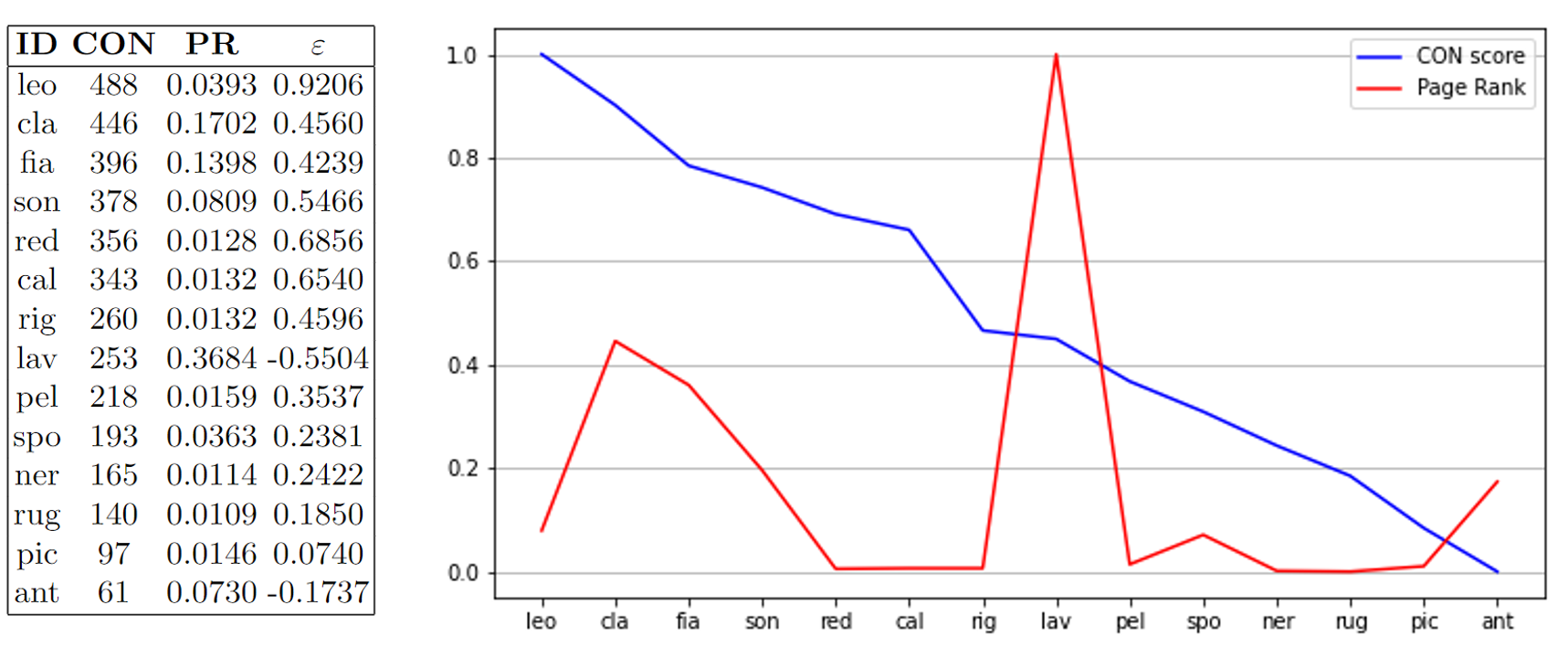}
\caption{CON score versus PageRank in the Bonanni2007-2 dominance network. Nodes such as leo, fia, and cla represent a population of mongrel dogs, and edges correspond to an observed dominance behavior. The table lists the CON scores, PageRank (PR), and low-key leader strengths ($\varepsilon$). The histogram depicts normalized CON scores versus PageRank.}\label{bonfig}
\end{figure}

\begin{figure}[tpbh!]
\centering
\includegraphics[scale=0.26]{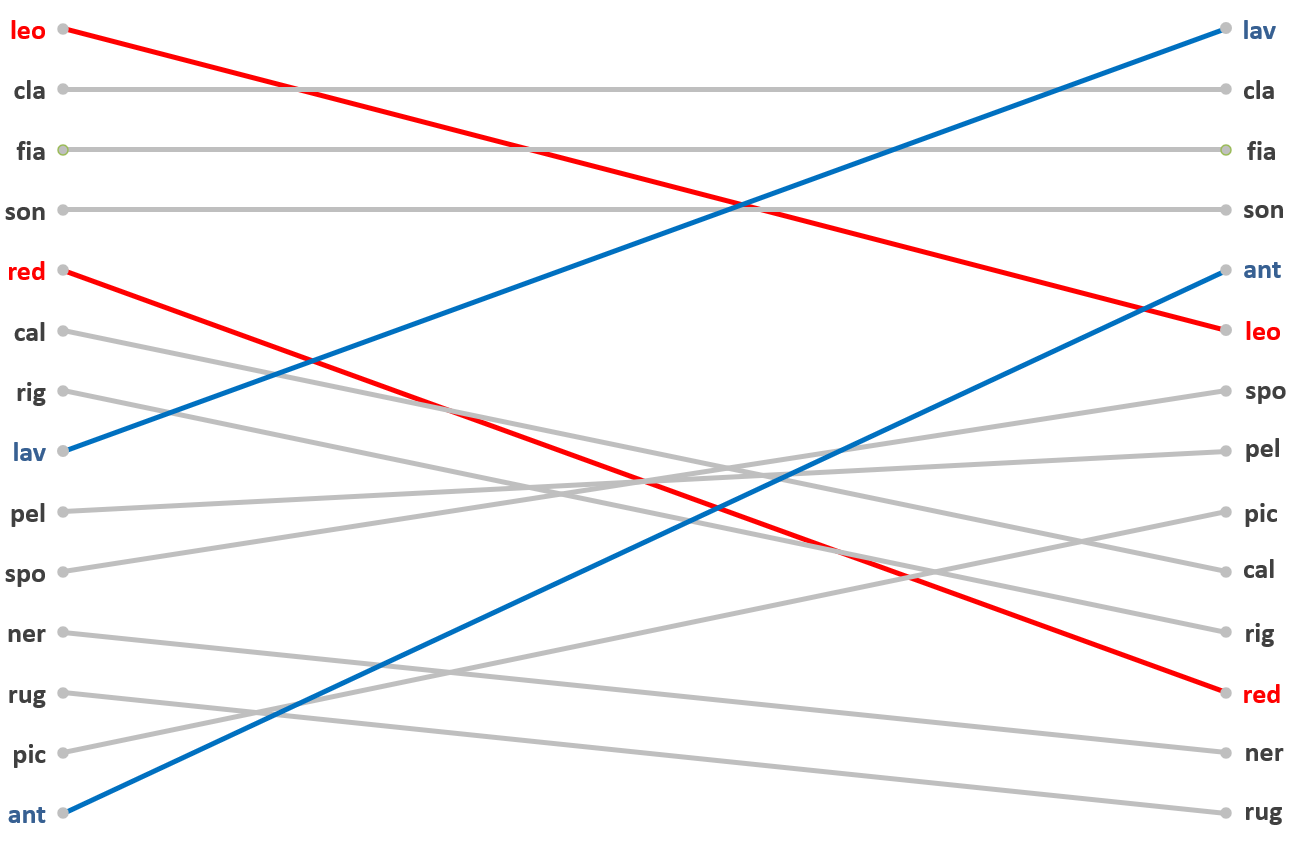}
\caption{A Slope Graph to compare the rankings via CON and PageRank in the Bonanni2007-2 dominance network. On the left, the top nodes 
via CON scores, while on the right, the top nodes via PageRank on
the reversed-edge network. Nodes are labeled in grey if the difference in rankings is less than five.
Nodes are labeled in red if the CON ranking is at least five places higher than the PageRank, and in blue
if the PageRank is at least five places higher than the CON score.}\label{bonsl}
\end{figure}

\begin{figure}[htpb!]
\centering
\includegraphics[scale=0.24]{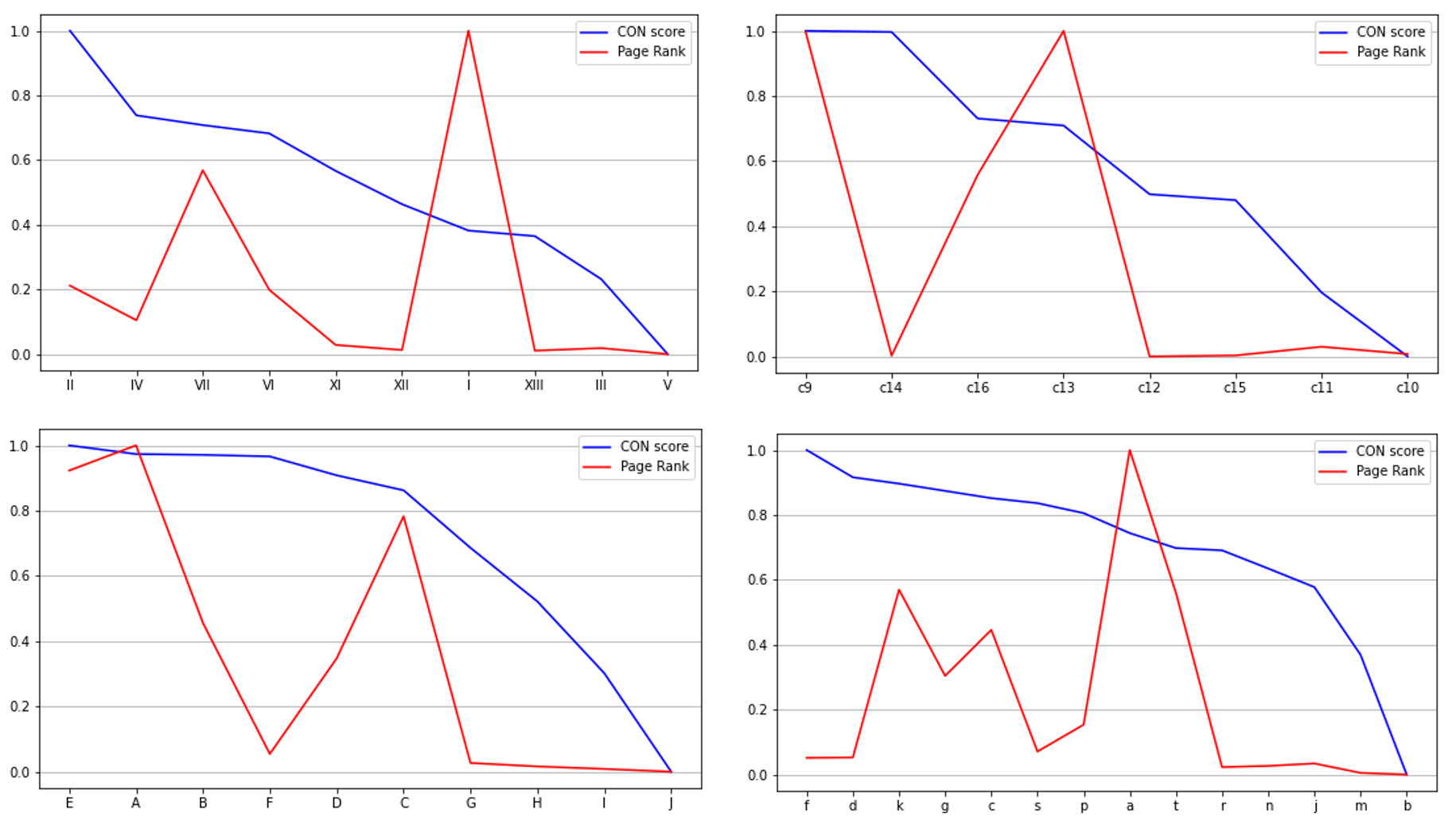}
\caption{CON scores versus PageRank in four animal dominance populations. From top left and clockwise, we list the data set name, predicted low-key leader, and relevant citation: Allee1954-3, II, \cite{allee}; Poisbleau2005-1a, C14, \cite{pois}; deWaal1977-1, f, \cite{dewall}; and Watt1986-1a, F, \cite{watt}.}\label{lollibon}
\end{figure}

See Figure~\ref{bonfig} for a comparison of centrality scores in the Bonanni2007-2 network. We seriate the animals via the difference in their CON score and PageRank from the highest value to the lowest. A slope graph representation of the data is provided in Figure~\ref{bonsl} and compares the rankings with CON scores and PageRank. The dog leo emerges as having the top CON score and highest difference between their CON scores and PageRank, with low-key leader strength $0.9206.$ We therefore determine that leo is the low-key leader in the Bonanni2007-2 network. 

We considered the low-key leader strengths of nodes in all the 172 animal dominance networks and detected LKLs in 155 or 90.12\% of them using a low-key leader strength of $\varepsilon = 0.5$. If we let choose $\varepsilon$ to be at least $0.4,$ then 95.35\% of them contain a low-key leader. We think the prevalence of LKLs in these dominance networks provides support for our hypothesis. Figure~2 compares CON scores and PageRank in four other animal dominance populations \cite{allee,dewall,pois,watt}. As referenced at the beginning of the section, see \url{https://github.com/jkapusin/Low-Key-Leaders} for a list of all the detected LKLs. 

\subsection{Trade Networks}

UN Comtrade \cite{COM} is a statistical database storing international trading information between nations that is organized by the United Nations Statistics Division. There are over 170 nations reporting their annual international trading data in the database. 

\begin{figure}[h!]
\centering
\includegraphics[scale=0.23]{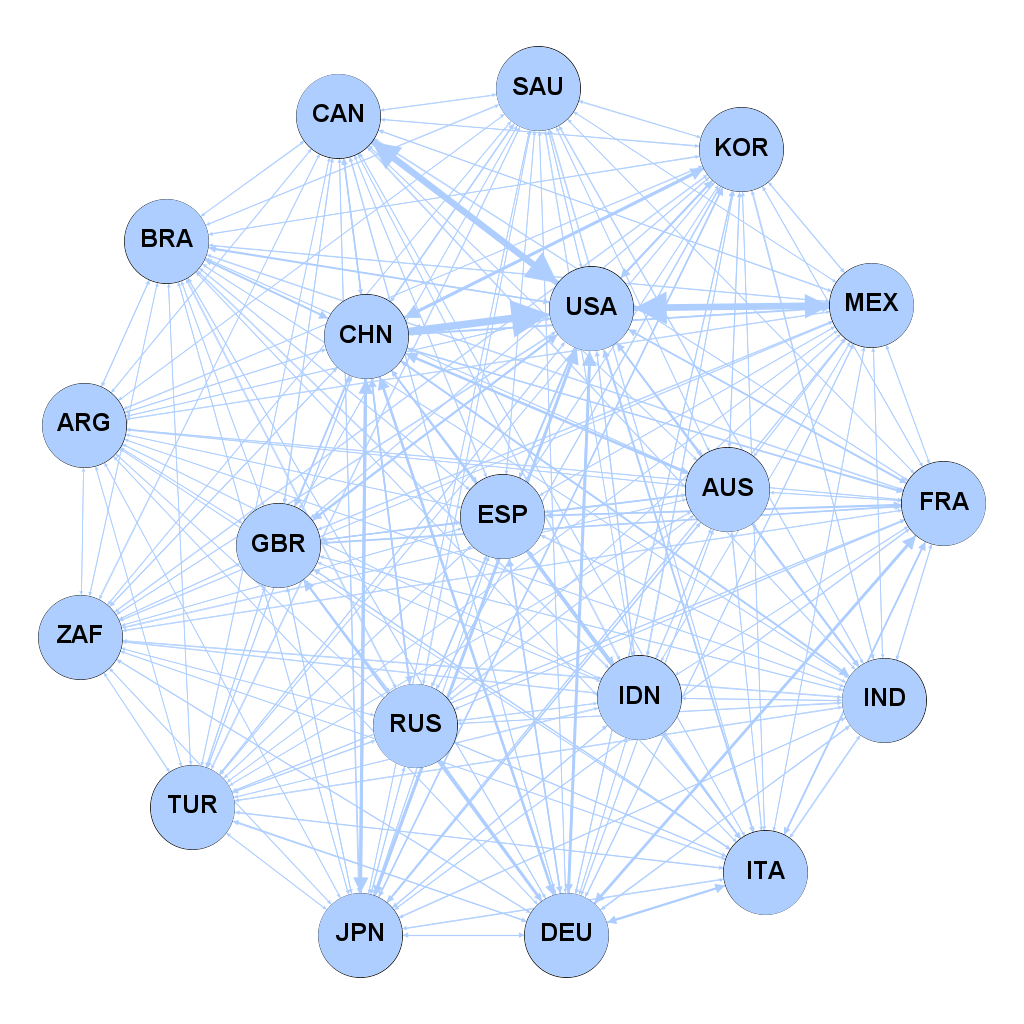}
\caption{The G20 trade network.}\label{g20fig}
\end{figure}

\begin{figure}[htpb!]
\centering
\includegraphics[scale=0.3]{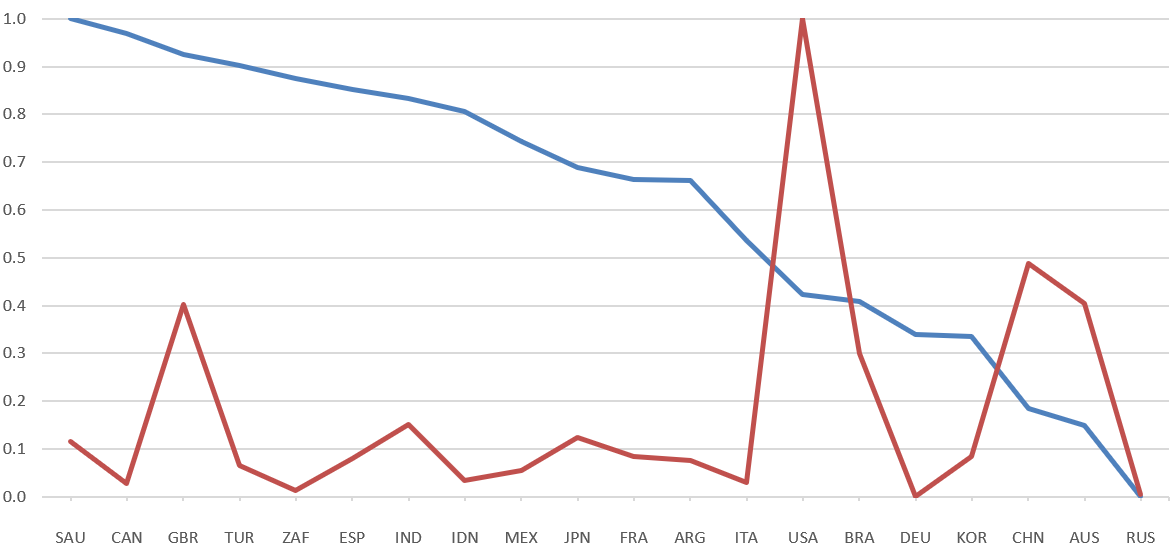}
\caption{CON score (blue) versus PageRank (red) for the trade deficit network of G20 nations. Nodes are nations denoted by three-letter codes such as CAN, USA, and CHN.}\label{weighted_com1}
\end{figure}

We extract trading data of the 19 nations within the G20 and Spain from 2019. An edge directed inward to the reporter nation represents importation, while edges directed outwards represent exportation. Hence, nations with larger in-degree than out-degree have larger trading deficits; a trading deficit between nations may be viewed as form of dominance or adversarial relationship. Trading volumes are considered as the weights of the edges in the weighted graphs. See Figure~\ref{g20fig}.

\begin{figure}[tpbh!]
\centering
\includegraphics[scale=0.33]{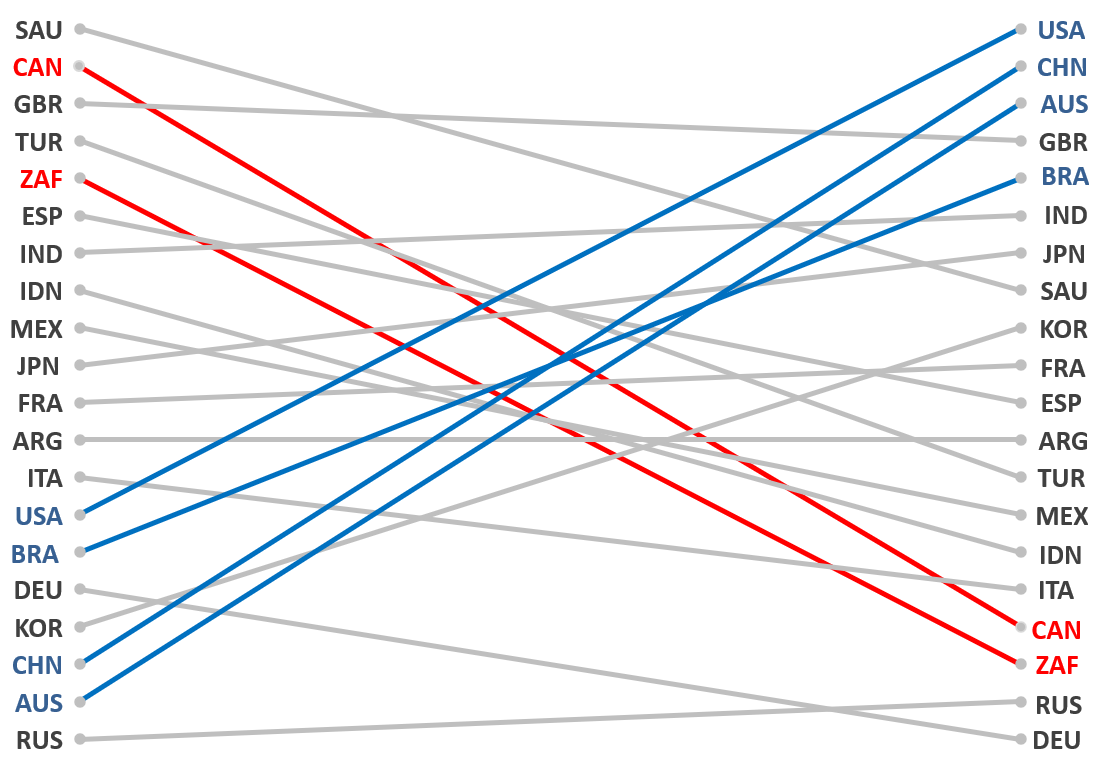}
\caption{A Slope Graph to compare the rankings via CON and PageRank in the G20 trade deficit network. On the left, the top nodes 
via CON scores, while on the right, the top nodes via PageRank. Nodes are labeled in grey if the difference in rankings is less than ten.
Nodes are labeled in red if the CON ranking is at least ten places higher than the PageRank, and in blue
if the PageRank is at least ten places higher than the CON ranking.}\label{g20sl}
\end{figure}

Figure~\ref{weighted_com1} is the histogram of the weighted networks of the CON score versus PageRank of the trading networks. Figure~\ref{g20sl} is the corresponding slope graph. Among these nations, Canada (CAN) emerges as a low-key leader with low-key leader strength 0.9401. These results support that view that while Canada does not have the highest trade in the G20, it plays an influential secondary role in shaping international trade dynamics among the higher income nations.

\subsection{Bitcoin Trust Networks}

Our final data set consists of a graph with a much larger number of nodes and edges than the dominance networks and trading networks. Users trading the cryptocurrency Bitcoin may anonymously rate others on their trustworthiness. The members of Bitcoin trust networks rate other members by assigning an integer from -10 (total distrust) to +10 (total trust); see \cite{Bit2}. We formed an adversarial network with nodes the users and edges corresponding to negative ratings; for example, if user $x$ rates user $y$ with -2, then we formed a directed edge $(x,y).$ The rating scores are considered as the weights in the weighted network. Data for Bitcoin trust networks was taken from \cite{snap}.

\begin{figure}[H]
\centering
\includegraphics[scale=0.23]{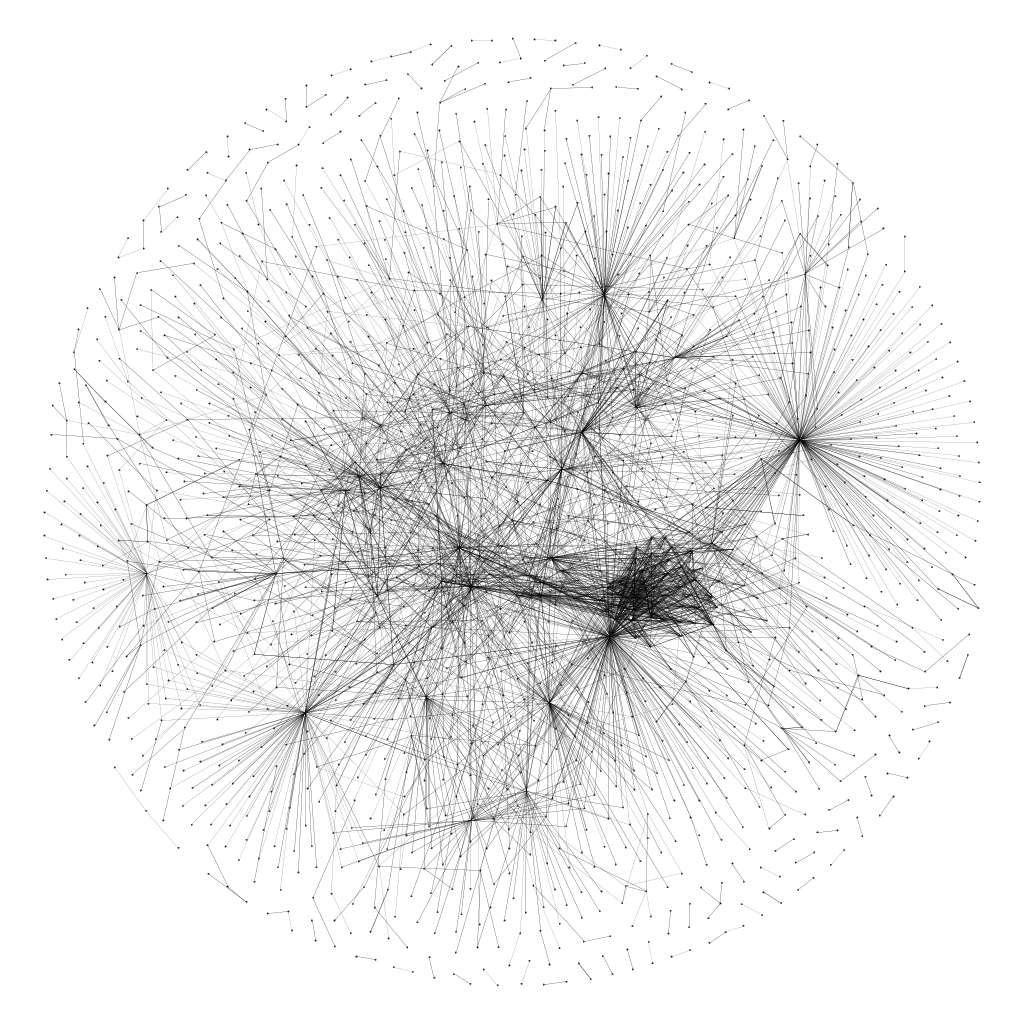}
\caption{A visualization of the Bitcoin Over-The-Counter (or OTC) trust network, where nodes are users and directed edges correspond to negative ratings between them.}\label{bit}
\end{figure}

\begin{figure}[H]
\centering
\includegraphics[width= 4.6 in]{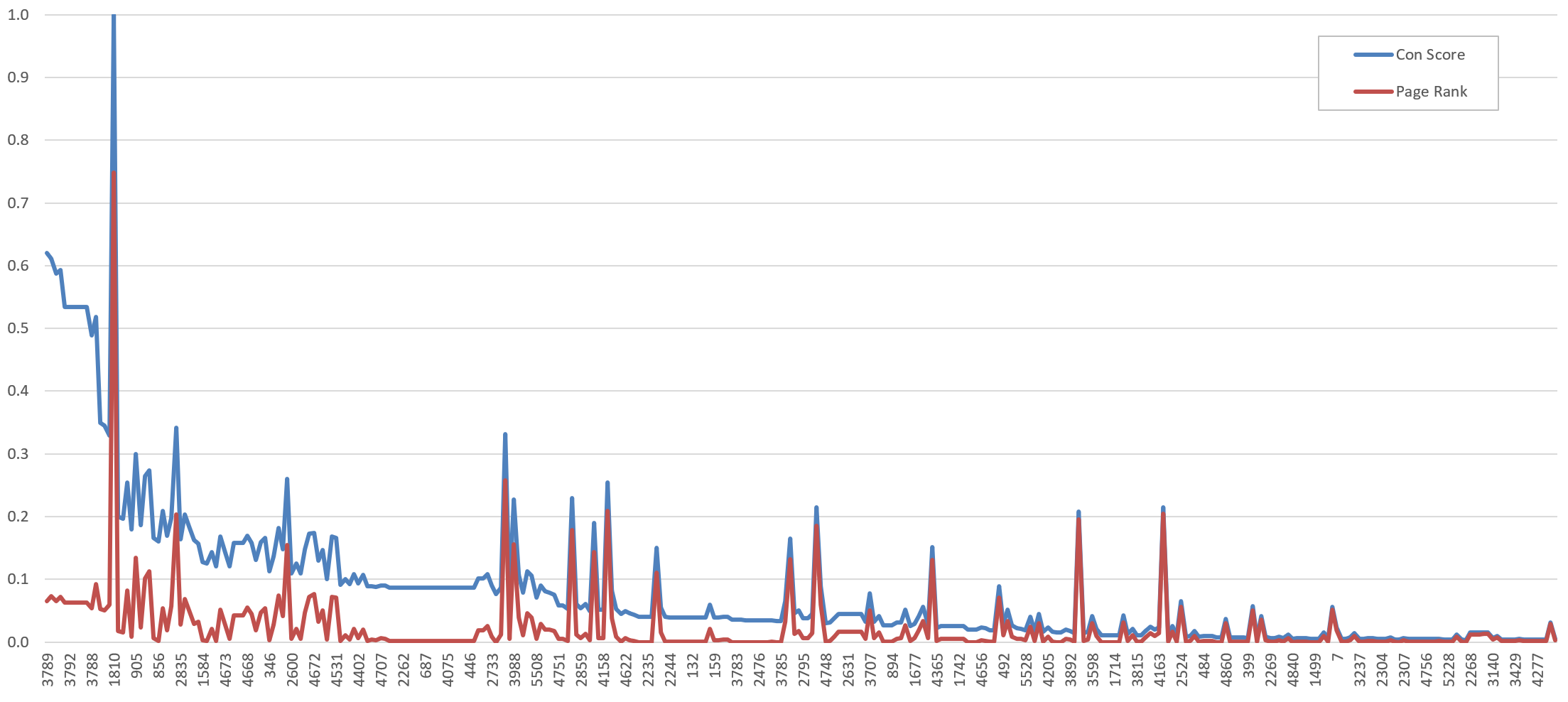}
\caption{CON (blue) and PageRank (red) scores in the OTC Bitcoin trust network. }\label{bit1}
\end{figure}

Figure~\ref{bit} provides a visualization of the OTC Bitcoin trust network with 5,882 nodes and 3,563 edges. As the number of users is large, we select the top 340 users by sorting the difference between CON score and PageRank of each user ranked from the highest to the lowest. Users outside this set of 340 had scores at or near zero and so were omitted. The histogram contrasting CON scores and PageRank is given in Figure~\ref{bit1}. The user with ID 3789 emerged as a LKL from our analysis, with low-key leader strength 0.5552. 


\section{Directed Ranking Model} 

Many models for complex networks were proposed over the last two decades involving various mechanisms such as preferential attachment and copying; see \cite{ab}, for example, for an early survey. We introduce a random directed graph generation model based on ranked-based attachment that simulates digraphs containing a low-key leader. In rank-based attachment models, the degree of a node is a function of their predetermined rank. An undirected rank-based attachment model was introduced in \cite{r12}. Such models are offline, in the sense that the number of nodes will not change over time. 

The \emph{directed ranking model} produces a sequence of digraphs $G_n$ with nodes $V_n =\{1,2,\ldots,n\} $, where $n\ge 1$ is an integer. The model has a fixed parameter and an adjustable parameter: the order of the digraph $n \in \mathbb{N}^+$ is fixed and the attachment strength $\alpha$ is chosen in $(0,1)$. Note that $n$ will not change over time. For each $v_i \in V_n$, it receives a label $l(n_i) \in \{1,2,\ldots,n\}$ chosen uniformly at random. Nodes are ranked based on their labels; that is, we denote $l$ as the label which node receives and $r$ as its corresponding rank. If $l(v_i) < l(v_j)$, then the nodes are ranked $r(v_i) > r(v_j)$. Each node has a unique rank and the node that receives label $1$ obtains the highest rank among all, while whichever node receives label $n$ has the the lowest rank.  For simplicity, we reorder the sequence of nodes and simply let $r(v_i)=i$ for all choices of $i$.

Edges in the model are added according to the attachment strength $\alpha$. They are generated by following the random process: for each distinct pair of nodes $v_i$ and $v_j$, the probability of generating a directed edge $(i,j)$ equals $$\mathbb{P}((i,j) \in E(G_n))=j^{-\alpha}.$$ We set $\alpha = \frac{1}{2}$ for simplicity. We next uniformly choose a node at random from the existing nodes, say $v_m$, that we call the \emph{copy node}. Let $v_r$ be the node with the highest out-degree. We deterministically add directed edges $(v_m,v_j)$, for every directed edge $(v_r,v_j)$. The copy node $v_m$ has high out-degree but note that its in-degree remains unchanged.

Note that for a node $v_i$, its in-degree $\mathrm{deg}^-(v_i)$ is the sum of $n-1$ independent Bernoulli trials with a predetermined probability based on the ranking scheme. The expected in-degree of a node $v_i$ with strength attachment $\alpha = \frac{1}{2}$ may be expressed as: 
\begin{eqnarray*}
\mathbb{E}(\mathrm{deg}^-(v_i)) &= &\sum_{j=1}^{n}i^{-\frac{1}{2}} \\
&=&i^{-\frac{1}{2}} \sum_{j=1}^{n} 1 \\
&=& ni^{-\frac{1}{2}}.
\end{eqnarray*}
Therefore, the expected in-degree of $v_i$ is one of $n,\frac{n}{\sqrt{2}}, \frac{n}{\sqrt{3}}, \dots,\sqrt{n}$. 
\begin{figure}
\centering
\begin{subfigure}{.5\textwidth}
  \centering
  \includegraphics[width=2.4in]{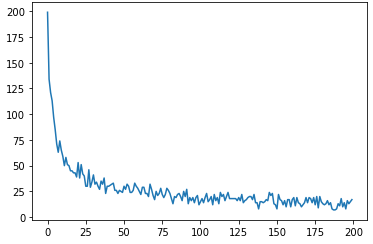}
  \caption{The in-degree distribution of $G_{200}$.}
\end{subfigure}%
\begin{subfigure}{.5\textwidth}
  \centering
  \includegraphics[width=2.4in]{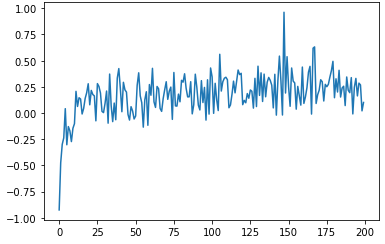}
  \caption{$\varepsilon$ values of $G_{200}$.}
  \label{fig:sub2}
\end{subfigure}
\caption{Simulation of the directed ranking model with 200 nodes.}\label{indeg}
\label{fig:test}
\end{figure}

We consider an example simulating a digraph with $n=200$ using the directed ranking model. As shown in Figure~\ref{indeg} (a), the in-degree distribution possesses a heavy tail. We determined the CON scores and PageRank in our simulated digraph. As depicted in Figure~\ref{indeg} (b), the copy node $v_{148}$ is the low-key leader as it has the largest low-key leader strength equaling 0.9354. 

\section{Discussion and Future Work}

We introduced a new type of node in adversarial networks using centrality measures. A low-key leader (or LKL) corresponds to a node that has a relatively high CON score, but relatively low PageRank. We asserted that LKLs typically exist in adversarial networks. To validate the assertion in real-world networks, we analyzed three different types of adversarial networks: animal dominance networks in 172 animal populations, trading networks between G20 nations, and Bitcoin trust networks. The analysis of the contrasting CON scores and PageRank in the three types of data sets supported presence of LKLs in trade networks, Bitcoin trust networks, and in over 90\% of the considered animal dominance networks. We introduced the directed ranking model that generates with high probability digraphs that have an expected power law in-degree distribution, and also possess low-key leaders.

While we provided evidence for our hypothesis on LKLs using three different types of networked data sets, we may consider networks in other knowledge domains to further validate their presence. We will consider a more rigorous analysis of the directed ranking model in future work, exploring concentration results on the in-degree distribution, as well as small world and spectral properties. An on-line version of the directed ranking model, where new nodes are introduced over time, will be considered in the full version of the paper. Finally, while we predicted the existence of the low-key leaders in adversarial networks, we do not posit in this work why they exist. The underlying mechanism as to why LKLs appear to be prevalent in adversarial networks remains open.

\end{document}